\documentclass[aps,prl,twocolumn,superscriptaddress,groupedaddress,preprintnumbers]{revtex4-1} 
\usepackage{graphicx}  
\usepackage{dcolumn}   
\usepackage{bm}        
\usepackage{amssymb}   
\usepackage{amsmath}
\usepackage{mathtools}
\usepackage{cases}
\usepackage{mathrsfs}
\usepackage{color}
\usepackage{enumitem}
\makeatletter
\newcommand*\bigcdot{\mathpalette\bigcdot@{.5}}
\newcommand*\bigcdot@[2]{\mathbin{\vcenter{\hbox{\scalebox{#2}{$\m@th#1\bullet$}}}}}
\makeatother
\newcommand{\mpl}{M_{\text{Pl}}}

\newcommand{\calP}{{\cal P}}

\newcommand{\be}{\begin{equation}}
\newcommand{\ee}{\end{equation}}
\newcommand{\bea}{\begin{eqnarray}}
\newcommand{\eea}{\end{eqnarray}}
\newcommand{\nn}{\nonumber}
\newcommand{\bx}{\mathbf{x}}
\newcommand{\bk}{\mathbf{k}}

\newcommand{\vf}{\varphi}

\newcommand*{\dif}{\mathop{}\!\mathrm{d}}
\begin{document}
	\preprint{IPMU21-0075}
	\preprint{YITP-21-131}
	\title{Primordial Black Hole Formation in Non-Minimal Curvaton Scenario}
	\author{Shi Pi${}^{a,b,c}$ and Misao Sasaki${}^{c,d,e}$\\
		\it
		$^{a}$ CAS Key Laboratory of Theoretical Physics, Institute of Theoretical Physics,\\
		Chinese Academy of Sciences, Beijing 100190, China \\
		$^{b}$ Center for High Energy Physics, Peking University, Beijing 100871, China\\
		$^{c}$ Kavli Institute for the Physics and Mathematics of the Universe (WPI), Chiba 277-8583, Japan\\		
		$^{d}$ Yukawa Institute for Theoretical Physics, Kyoto University, Kyoto 606-8502, Japan\\
		$^{e}$ Leung Center for Cosmology and Particle Astrophysics,\\National Taiwan University, Taipei 10617}
	\date{\today}
\begin{abstract}
In the curvaton scenario, the curvature perturbation is generated after inflation at the curvaton decay, which may have a prominent non-Gaussian effect. For a model with a non-trivial kinetic term, an enhanced curvature perturbation on a small scale can be realized, which can lead to copious production of primordial black holes (PBHs) and induce secondary gravitational waves (GWs). Using the probability distribution function (PDF) which takes full nonlinear effects into account, we calculate the PBH formation. We find that under the assumption that thus formed PBHs would not overclose the universe, the non-Gaussianity of the curvature perturbation can be well approximated by the local quadratic form, which can be used to calculate the induced GWs. In this model the limit of large non-Gaussianity can be reached when the curvaton energy fraction $r$ is small at the moment of curvaton decay. We also show that in the $r\to1$ limit the PDF is similar to that of ultraslow-roll inflation.
	\end{abstract}
	\maketitle

\textit{Introduction.}---The discovery of gravitational waves (GWs) by LIGO in 2015 marked the dawn of GW cosmology \cite{Abbott:2016blz}. Since then about 90 GW events have been detected~\cite{LIGOScientific:2021djp}.
Besides the transient GWs emitted from mergers, the stochastic GW background is an important scientific target of LIGO/VIRGO/KAGRA~\cite{KAGRA:2021kbb} and the space-borne interferometers like LISA~\cite{Barausse:2020rsu}, Taiji~\cite{Hu:2017mde,Ruan:2018tsw}, and TianQin~\cite{TianQin:2015yph}. 
Among all the sources of stochastic GWs, secondary GWs induced by a peaked primordial curvature perturbation~\cite{Matarrese:1992rp,Matarrese:1993zf,Matarrese:1997ay,Noh:2004bc,Carbone:2004iv,Nakamura:2004rm,Ananda:2006af,Osano:2006ew,Baumann:2007zm} have recently attracted much attention \cite{Assadullahi:2009jc,Alabidi:2012ex,Alabidi:2013lya,Biagetti:2014asa,Inomata:2016rbd,Nakama:2016gzw,Orlofsky:2016vbd,Gong:2017qlj,Garcia-Bellido:2017aan,Kohri:2018awv,Cai:2018dig,Bartolo:2018rku,Unal:2018yaa,Adshead:2021hnm,Yuan:2020iwf,Pi:2020otn,Domenech:2021ztg,Liu:2021jnw} as they imply the existence of abundant primordial black holes (PBHs)~\cite{Saito:2008jc,Saito:2009jt,Bugaev:2009zh,Bugaev:2010bb}. PBHs can play a number of important roles in cosmology. They may provide the seeds for galaxy formation~\cite{Bean:2002kx,Kawasaki:2012kn,Nakama:2017xvq,Carr:2018rid,Nakama:2019htb,Carr:2020erq,Atal:2020yic}, may account for a population of the LIGO-Virgo 
events~\cite{Bird:2016dcv,Clesse:2016vqa,Sasaki:2016jop,Blinnikov:2016bxu,Ali-Haimoud:2016mbv,Zumalacarregui:2017qqd,Garcia-Bellido:2017imq,Wong:2020yig,Hutsi:2020sol,Kimura:2021sqz,Franciolini:2021tla}, the ultra-short timescale microlensing events~\cite{2017Natur.548..183M,Niikura:2019kqi}, the planet 9~\cite{Scholtz:2019csj}, hot spot for baryogenesis~\cite{Byrnes:2018clq,Carr:2019hud,Garcia-Bellido:2019vlf,Carr:2019kxo,DeLuca:2021oer}, and the cold dark matter~\cite{Carr:2016drx,Niikura:2017zjd,Katz:2018zrn,Katz:2018zrn,Montero-Camacho:2019jte,Sugiyama:2019dgt,Laha:2019ssq,DeRocco:2019fjq}. 
In the last case, the curvature perturbation on uniform density slices $\zeta$ should be enhanced to $\mathcal O(0.1)$ at $k_*\sim10^{13}~\text{Mpc}^{-1}$, which can be realized by, for instance, 
ultra-slow-roll single-field inflation \cite{Yokoyama:1998pt,Garcia-Bellido:2016dkw,Cheng:2016qzb,Garcia-Bellido:2017mdw,Cheng:2018yyr,Dalianis:2018frf,Tada:2019amh,Xu:2019bdp,Mishra:2019pzq,Bhaumik:2019tvl,Liu:2020oqe,Atal:2019erb,Fu:2020lob,Vennin:2020kng,Ragavendra:2020sop,Gao:2021dfi,Pattison:2021oen,Ng:2021hll}, 
modified gravity~\cite{Kannike:2017bxn,Pi:2017gih,Gao:2018pvq,Cheong:2019vzl,Cheong:2020rao,Fu:2019ttf,Dalianis:2019vit,Lin:2020goi,Fu:2019vqc,Aldabergenov:2020bpt,Aldabergenov:2020yok,Yi:2020cut,Gao:2020tsa,Dalianis:2020cla,Kawai:2021edk},
multi-field inflation with a flat potential~\cite{GarciaBellido:1996qt,Kawasaki:1997ju,Frampton:2010sw,Clesse:2015wea,Inomata:2017okj,Inomata:2017vxo,Espinosa:2017sgp,Kawasaki:2019hvt,Palma:2020ejf,Fumagalli:2020adf,Braglia:2020eai,Anguelova:2020nzl,Romano:2020gtn,Gundhi:2020zvb,Gundhi:2020kzm}, resonance~\cite{Cai:2018tuh,Chen:2020uhe,Cai:2019jah,Cai:2019bmk,Cai:2021wzd,Cai:2021yvq}, oscillons~\cite{Cotner:2016cvr,Cotner:2017tir,Cotner:2018vug,Cotner:2019ykd}, etc.

In this paper we propose a non-minimal curvaton model that can produce such an enhanced, peaked curvature perturbation
at $k=k_*$. In the curvaton scenario, the primordial curvature perturbation is produced by the curvaton field perturbation that was isocurvature during inflation but turns into the curvature perturbation as the curvaton starts to dominate the universe \cite{Moroi:2001ct,Enqvist:2001zp,Lyth:2002my}.  A simple mechanism to generate a peaked spectrum is to introduce a non-trivial field metric which suppresses the curvaton kinetic term around $k=k_*$. The curvature perturbation produced by curvaton is intrinsically non-Gaussian \cite{Bartolo:2003jx,Enqvist:2005pg,Sasaki:2006kq}. Using the probability distribution function which takes full nonlinear effects into account, we calculate the PBH formation. We find that under the assumption that the mean square value of $\zeta$ is small, $\left\langle\zeta^2\right\rangle\lesssim0.1$, which is justified \textit{a posteriori} by the condition that the produced PBHs would not overclose the universe, 
the resultant fully nonlinear curvature perturbation can be well described by the quadratic local non-Gaussianity, $\zeta=\zeta_g+F_\text{NL}(\zeta_g^2-\langle\zeta_g^2\rangle)$, where $F_\text{NL}$ is inversely proportional to the curvaton energy fraction at its decay moment, which can freely be very large. 


\textit{Non-minimal Curvaton.}--- We consider a two-field theory with a non-trivial field metric,
\be
\mathscr{L}=-\frac12\big(\partial\varphi\big)^2-V(\varphi)-\frac12f(\varphi)^2\big(\partial\chi\big)^2-\frac12m_\chi^2\chi^2,
\ee
where $\varphi$ is the inflaton, $V(\varphi)$ is its potential, and $\chi$ is the curvaton with mass $m_\chi$. 
Such a field metric may appear in dilatonic or axionic models or in modified gravity~\cite{Berkin:1991nm,Domenech:2015qoa}.
We assume single-field slow-roll inflation with the energy density dominated by $V(\varphi)$.
The background field equations are
\begin{gather}\label{eom:vf}
\ddot\varphi+3H\dot\varphi+V'(\varphi)=ff'\dot\chi^2\approx0,\\\label{eom:chi}
\ddot\chi+3H\dot\chi+2\frac{f'}{f}\dot\vf\dot\chi+\frac{m_\chi^2\chi}{f(\varphi)^2}=0,
\end{gather}
where $H\approx \sqrt{8\pi GV(\varphi)/3}$ is the Hubble parameter during inflation. The approximation in \eqref{eom:vf} holds because the curvaton is effectively ``frozon'' as we assume $m_\chi\ll H$. The equation of motion for the curvaton perturbation on the spatially-flat slicing, $\delta\chi$, is
\begin{align}\label{eom:deltachi}
&\ddot{\delta\chi}+\left(3H+2\frac{f'}{f}\dot\vf\right)\dot{\delta\chi}+\left[\frac{k^2}{a^2}+\frac{m_\chi^2}{f^2}\right]\delta\chi\approx0,
\end{align}
where we neglected the terms proportional to $\dot\chi$. Deep inside the horizon, the WKB solution to \eqref{eom:deltachi} for the initial conditions in the adiabatic vacuum  is
\be\label{deltachisub}
\quad\delta\chi=\frac{1}{\sqrt{2k}a f(\vf)}\exp\left(-ik\int\frac{\dif t}{a}\right).
\ee
On superhorizon scales, the $(k^2/a^2)$-term in \eqref{eom:deltachi} is negligible. So \eqref{eom:chi} and \eqref{eom:deltachi} have the same form, which gives $\delta\chi\propto\chi$, and 
\be
\frac{\delta\chi(t)}{\chi(t)}=\frac{\delta\chi(t_k)}{\chi(t_k)}\approx\frac{H(t_k)}{\sqrt{2k^3}f(t_k)\chi(t_k)},
\label{dchi/chi}
\ee
for $t>t_k$ where $t_k$ is the horizon crossing time determined by $k=a(t_k)H(t_k)$. 
Thus the power spectrum of $\delta\chi/\chi$ on the spatially-flat slicing is
\be\label{def:Pdelta1}
\calP_{\delta\chi/\chi}(k)=\frac{k^3}{2\pi^2}\left|\frac{\delta\chi_\bk}{\chi}\right|^2=\frac{1}{\chi(t_k)^2}\left(\frac{H(t_k)}{2\pi f(t_k)}\right)^2.
\ee
We assume there is a sharp dip in $f(\varphi)$ which results in a peak in $1/f(t_k)$ at $k=k_*$.  
Around the peak, $H(t_k)$ and $\chi(t_k)$ are only very slowly varying. Hence their $k$-dependence may be neglected. For $k$ sufficiently far away from $k_*$, the right-hand side of (\ref{def:Pdelta1}) is slowly varying in $k$, which results in an almost scale-invariant spectrum of $\delta\chi/\chi$.

After inflation, the universe is radiation-dominated with its energy density behaves as $\rho_\text{r}\propto a^{-4}$. After the Hubble expansion rate $H$ drops below $m_\chi$, the curvaton begins to oscillate, and the energy density $\rho_\chi$ starts to behave as $a^{-3}$. 
The curvature perturbation on the uniform-curvaton-density slice is \cite{Lyth:2004gb}
\be\label{zetachi}
\zeta_\chi(t,\mathbf{x})=\phi+\int^{\rho_\chi(t,\bx)}_{\rho_\chi(t)}\frac{\dif\widetilde\rho_\chi}{3\widetilde\rho_\chi}
=\phi+\frac13\ln\frac{\rho_\chi(t,\bx)}{\rho_\chi(t)},
\ee
where $\phi(t,\bx)$ and $\rho_\chi(t)$ are the curvature perturbation and the ``background'' energy density of $\chi$ respectively, in an arbitrary gauge of $\{t,\bx\}$. 
We have
\be
\rho_\chi=m_\chi^2\big(\bar\chi+\delta\chi\big)^2=\bar\rho_\chi(t)\left(1+\frac{\delta\chi}{\bar\chi}\right)^2.
\ee
where $\bar\chi(t)$ is a fiducial value of $\chi$, and $\bar\rho_\chi\equiv m_\chi^2\bar\chi^2$. 
In spatially-flat slicing, \eqref{zetachi} becomes $\zeta_\chi(t,\mathbf{x})=(1/3)\ln(1+\delta)^2\equiv(1/3)\ln A$, 
where $\delta\equiv\delta\chi/\bar\chi$ is the curvaton contrast in the spatially-flat slicing, and $A=e^{3\zeta_\chi}$. In uniform-total-density slicing, \eqref{zetachi} becomes 
$\zeta_\chi=\zeta_\text{b}+(1/3)\ln\rho_\chi^\text{(u)}/\bar\rho_\chi^\text{(u)}$,  
which gives $\rho_\chi^\text{(u)}=\bar\rho^\text{(u)}_\chi e^{3(\zeta_\chi-\zeta_\text{b})}=A\bar\rho^\text{(u)}_\chi e^{-3\zeta_\text{b}}$. 
$\zeta_\text{b}$ is the ``bare'' curvature perturbation defined by choosing the time slicing on the right hand side of \eqref{zetachi} to be of uniform total-density, which has nonzero mean value and will be renormalized to $\zeta\equiv\zeta_\text{b}-\langle\zeta_\text{b}\rangle$ when we discuss the PBH formation later. Similarly for the radiation we have $\rho_\text{r}^\text{(u)}=\bar\rho^\text{(u)}_\text{r} e^{4(\zeta_\text{r}-\zeta_\text{b})}\approx\bar\rho^\text{(u)}_\text{r} e^{-4\zeta_\text{b}}$. The last step holds because we assume the curvature perturbation is mainly contributed by the curvaton, i.e., $\zeta_\text{r}\ll \zeta_\text{b}$.

We adopt the sudden-decay approximation to calculate the curvature perturbation, which is in good agreement with the numerical results~\cite{Malik:2006pm}. The curvaton decays when the Hubble rate $H$ equals to the decay rate $\Gamma$ on the hypersurface of uniform-total-density,
\be
\rho_\text{r}^\text{(u)}(t_\text{dec},\mathbf{x})+\rho_\chi^\text{(u)}(t_\text{dec},\mathbf{x})=\bar\rho^\text{(u)}(t_\text{dec})=3\mpl^2\Gamma^2.
\ee
Then at $t=t_\text{dec}$ we have
\be\label{zetaeqn}
e^{4\zeta_\text{b}}-\frac{4r}{3+r}Ae^{\zeta_\text{b}}+\frac{3r-3}{3+r}=0,
\ee
where
\begin{align}
r&\equiv\frac{3\Omega_{\chi,\text{dec}}}{4-\Omega_{\chi,\text{dex}}}=\left.\frac{3\bar\rho^\text{(u)}_\chi}{3\bar\rho^\text{(u)}_\chi+4\bar\rho^\text{(u)}_\text{r}}\right|_\text{dec},\\\label{def:A}
A&\equiv e^{3\zeta_\chi}
=\left(1+\delta\right)^2.
\end{align}
After the decay, $t>t_\text{dec}$, $\zeta_\text{b}$ is conserved until it reenters the Hubble horizon. Eq.~\eqref{zetaeqn} is a fourth order algebraic equation for $e^{\zeta_\text{b}}$, with the real positive solution given by \cite{Sasaki:2006kq}
\be\label{sol:zeta}
e^{\zeta_\text{b}}=K^{1/2}\frac{1+\sqrt{ArK^{-3/2}-1}}{(3+r)^{1/3}},
\ee
where
\begin{align}
K&=\frac12\left[P^{1/3}+(r-1)(3+r)^{1/3}P^{-1/3}\right],\\
P&=(Ar)^2+\sqrt{(Ar)^4+(3+r)(1-r)^3}.
\end{align}
As we commented, ``physical'' curvature perturbation should be defined as $\zeta\equiv\zeta_\text{b}-\langle\zeta_\text{b}\rangle$. 

When $\zeta_\text{b}$ is small, \eqref{sol:zeta} can be expanded as
\begin{align}\label{zetaseries}
\zeta_\text{b}&=\frac{r}{3} (A-1) -\frac{r^2(r+2)}{18} (A-1)^2+\cdots.
\end{align}
To the second order $\zeta$ has the familiar form of the quadratic local non-Gaussianity,
\be\label{localNG}
\zeta\approx\zeta_g+F_\text{NL}\left(\zeta_g^2-\left\langle\zeta_g^2\right\rangle\right),
\ee
with $\zeta_g=(2/3)r\delta$ and $F_\text{NL}=3/(4r)=\Omega_{\chi,\text{dec}}^{-1}-1/4$. We will see that \eqref{localNG} is a good approximation if we require that PBHs do not overclose the universe. 
 
From \eqref{zetaeqn} and \eqref{def:A}, we have 
\be\label{delta-zeta}
\delta
=-1\pm\sqrt{\frac{3+r}{4r}e^{3(\zeta+\langle\zeta_\text{b}\rangle)}+\frac{3r-3}{4r}e^{-\zeta-\langle\zeta_\text{b}\rangle}},
\ee
where $\delta\equiv\delta\chi/\bar\chi$. The ensemble average $\langle\zeta_\text{b}\rangle$ is calculated by $\langle\zeta_\text{b}\rangle=\int^\infty_{-\infty}\zeta_\text{b}P_\delta[\delta;R]\dif\delta$, where $P_\delta[\delta;R]$ is the Gaussian probability distribution function (PDF) of $\delta$,
\be\label{PDFdelta}
P_\delta[\delta;R]=\frac{1}{\sqrt{2\pi \sigma_{\delta}^2(R)}}\exp\left(-\frac{\delta^2}{2\sigma_{\delta}^2(R)}\right).
\ee
$\sigma_\delta^2(R)$ is the variance of $\delta$ smoothed on scale $R$, i.e., $\langle\delta^2\rangle=\int^\infty_{-\infty}\delta^2P_\delta[\delta;R]\dif\delta=\sigma_\delta^2(R)$. Therefore, 
the leading order of \eqref{zetaseries}, $\zeta_\text{b}\approx(r/3)\left(2\delta+\delta^2\right)$, gives $\langle\zeta_\text{b}\rangle\approx(r/3)\sigma_\delta^2(R)$, which can be checked \textit{a posteriori}.

In momentum space, $\sigma_\delta(R)$ is connected to the power spectrum of $\delta=\delta\chi/\bar\chi$ in \eqref{def:Pdelta1} as 
\begin{align}\nn
\sigma_{\delta}^2(R)&=\int^\infty_0\frac{\dif k}{k}\widetilde W^2(k;R)\mathcal{P}_{\delta}(k)\\\label{deltachi(R)} 
&\approx\int\left(\frac{H(t_k)}{2\pi\bar\chi(t_k)}\right)^2\frac{\widetilde W^2(k;R)}{f(t_k)^2}\frac{\dif k}{k}.
\end{align}
$\widetilde W^2(k;R)$ is the Fourier transform of a window function in real space, used for smoothing.
There are some ambiguities in the choice of window functions~\cite{Young:2019osy,Tokeshi:2020tjq}. For concreteness and simplicity, we choose the Gaussian window function, whose Fourier transform is $\widetilde W(k;R)=\exp\left(-k^2R^2/2\right)$. As mentioned before, $H$ and $\chi$ depend only weakly on $k$, while $f(\vf)$ has a sharp dip at $k_*$.
A dip in $f$ gives a peak in $1/f^2$ in \eqref{deltachi(R)}. So the power spectrum $\calP_\delta$ in \eqref{deltachi(R)} can be modeled by a lognormal function with the dimensionless width $\Sigma$,
\be\label{def:Pdelta}
\calP_\delta(k)=\frac{\sigma_0^2}{\sqrt{2\pi}\Sigma}\exp\left(-\frac{\ln^2(k/k_*)}{2\Sigma^2}\right).
\ee
We have neglected the near-scale-invariant component which is much smaller than the peak component. For simplicity, we focus on a narrow peak, $\Sigma\lesssim0.1$. Then the smoothed variance of the curvaton contrast in \eqref{deltachi(R)} is $\sigma^2_{\delta}(R)\approx\sigma_0^2\exp\left(-k_*^2R^2\right)$,
where $\sigma_0^2\equiv\sigma_\delta^2(R\to0)=\sqrt{2\pi}\Sigma\left[H_*/(2\pi\bar\chi_*f_*)\right]^2$ is the total power of $\mathcal P_\delta$.

\begin{figure}[htbp]
\begin{center}
\includegraphics[width=0.45\textwidth]{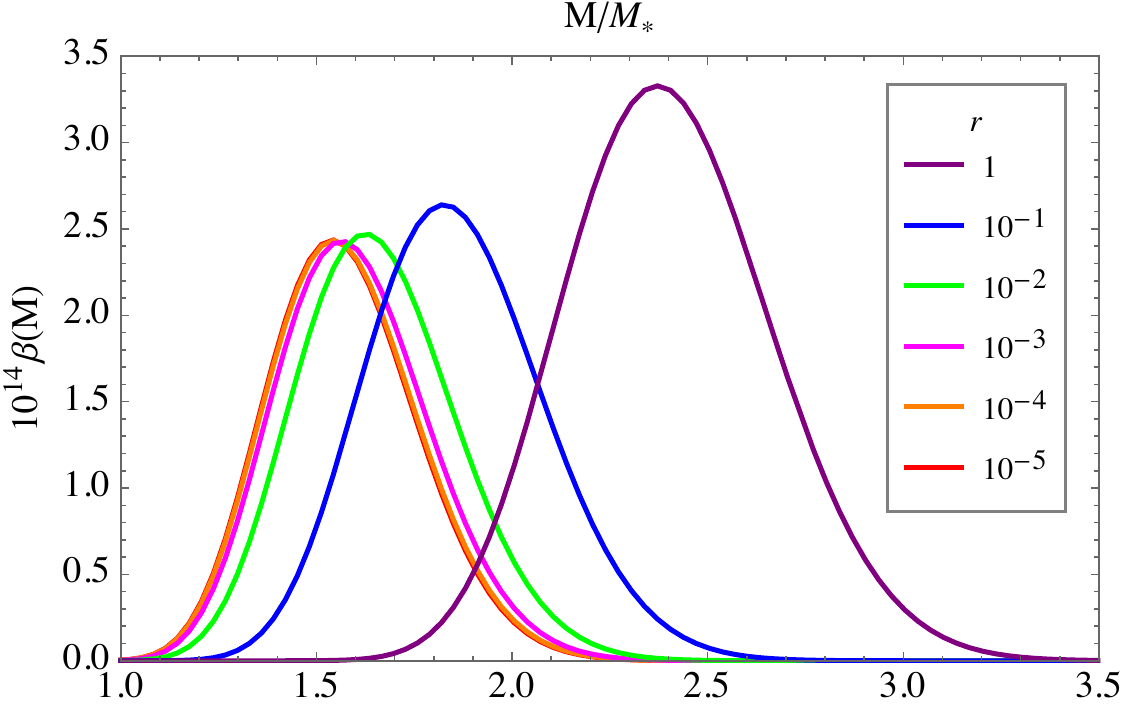}
\includegraphics[width=0.45\textwidth]{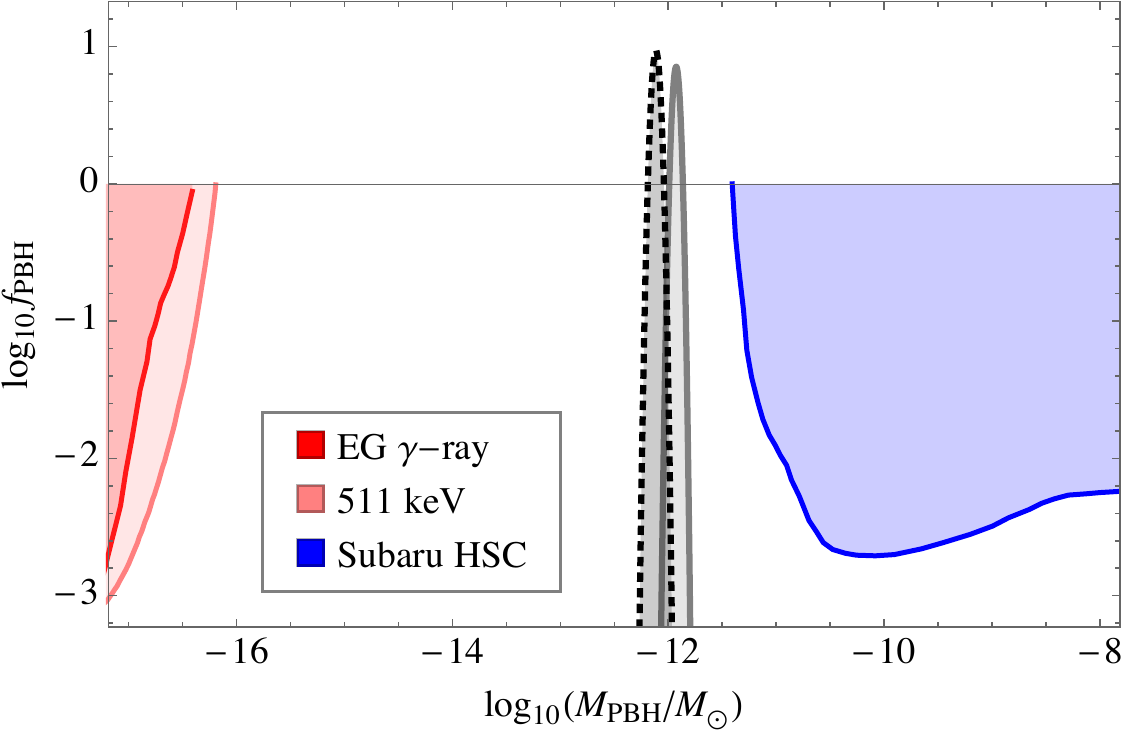}
\caption{
The top panel shows the PBH abundance at formation, $\beta(M)$, given by \eqref{beta-R}. The amplitude is normalized such that the PBHs constitute all the dark matter for $M_*=10^{-12}M_\odot$. 
The bottom panel shows the corresponding PBH mass function, $f_\text{PBH}(M)$. The solid and dashed curves are the cases of $r=1$ and $r=10^{-5}$, respectively. The observational constraints are from extra-galactic $\gamma$-rays  \cite{Carr:2020gox,Laha:2020ivk,Laha:2020vhg}, the 511 keV line from galactic center \cite{DeRocco:2019fjq,Laha:2019ssq,Dasgupta:2019cae,Ray:2021mxu}, and Subaru HSC~\cite{Niikura:2017zjd,Sugiyama:2019dgt}. Note that $f_\text{PBH}(M)$ exceeds unity because of the narrowness of the width.
}\label{f:beta-M}
\end{center}
\end{figure}

\vspace{1em}
\textit{PBHs and Induced GWs.}---Large density contrast may cause the PBH formation at the horizon reentry. The density contrast on the comoving slicing is 
\be\label{poisson}
\Delta\equiv\frac{\delta\rho^\text{(c)}}{\bar\rho^\text{(c)}}=-\frac{2\nabla^2\Phi}{3(Ha)^2}\approx-\frac{4\nabla^2\zeta}{9(Ha)^2},
\ee
where $\Phi$ and $\zeta$ are the curvature perturbations in the longitudinal and uniform-total-density slicings, respectively, which are related to each other as $\Phi=(2/3)\zeta$ on superhorizon scales in the radiation-dominated era~\cite{Kodama:1984ziu}.
The difficulty is that the operation is nonlocal when transforming the criterion for $\Delta$ to that for $\zeta$, $\zeta\propto \nabla^{-2}\Delta$.
Nevertheless, when the peak of $\zeta$ at $k_*$ is narrow, we may replace the gradient operator $\nabla$ by $-ik_*$.
Therefore, the inverse of \eqref{poisson} can be converted to an algebraic relation, $\zeta=9\Delta/(4k_*^2R^2)$, where $R=1/(Ha)$ is the comoving horizon scale, which is also the smoothing scale. From \eqref{delta-zeta} with $\langle\zeta_\text{b}\rangle\approx(r/3)\sigma_\delta^2(R)$, we have
\begin{align}\nn
\delta_\pm&=-1\pm\left[\frac{3+r}{4r}\exp\left(\frac{27\Delta}{4(k_*R)^2}+r\sigma_0^2\exp(-k_*^2R^2)\right)\right.\\\label{delta-Delta}
&\left.+\frac{3r-3}{4r}\exp\left(-\frac{9\Delta}{4(k_*R)^2}-\frac r3\sigma_0^2\exp(-k_*^2R^2)\right)\right]^{1/2}.
\end{align}

The PDF of $\Delta$ is now related to that of the curvaton contrast $\delta$ as
\be\label{Pchain}
P_\Delta[\Delta]\dif\Delta=P_\zeta[\zeta]\dif\zeta=P_\delta[\delta]\dif\delta.
\ee
The PBH abundance can be calculated by integrating the smoothed PDF of $\Delta$ from its critical value $\Delta_\text{cr}$, \`{a} la Press-Schechter,
\begin{align}\nn
\beta(R)&=\int_{\Delta_\text{cr}}P_\Delta[\Delta]\dif\Delta
=\int^\infty_{\delta_\text{cr}^+}P_\delta[\delta]\dif\delta+\int_{-\infty}^{\delta_\text{cr}^-}P_\delta[\delta]\dif\delta\\\label{beta-R}
&=\frac{1}{2}\left[\text{erfc}\left(\frac{\delta_\text{cr}^+}{\sqrt2\sigma_\delta(R)}\right)+\text{erfc}\left(\frac{|\delta_\text{cr}^-|}{\sqrt2\sigma_\delta(R)}\right)\right]\!,
\end{align}
where $\text{erfc}(x)$ is the complementary error function and the critical values $\delta_\text{cr}^\pm$ are given by \eqref{delta-Delta} with $\Delta=\Delta_\text{cr}$. 
The critical comoving density contrast $\Delta_\text{cr}$ varies from 0.2 to 0.6 in the literature. In this paper we adopt the value of $\Delta_\text{cr}$ inferred from
the optimized criterion given in \cite{Yoo:2018kvb}, where
the peak theory is used to determine the critical peak value of $\zeta$, $\zeta_\text{cr}\approx0.53$ for a specific profile in real space. Translating this result
into the Press-Schechter language, the corresponding value is estimated as $\Delta_\text{cr}\approx0.23$. 

The $R$-dependence of $\beta(R)$ can be converted to the PBH mass $M$ dependence
by $k_*R=(M/M_*)^{1/2}$,
where $M_*$ is the PBH mass when $k_*$ reenters the horizon. Fig.~\ref{f:beta-M} shows $\beta(M)$ with fixed PBH abundance today. The width of the peak is $\Delta M/M\sim\mathcal O(0.1)$, independent of $r$ and $\sigma_0$, reflecting the the fact that the width of $\cal P_\delta$ is small, $\Sigma\lesssim\mathcal{O}(0.1)$.

\begin{figure}[htbp]
\begin{center}
\includegraphics[width=0.41\textwidth]{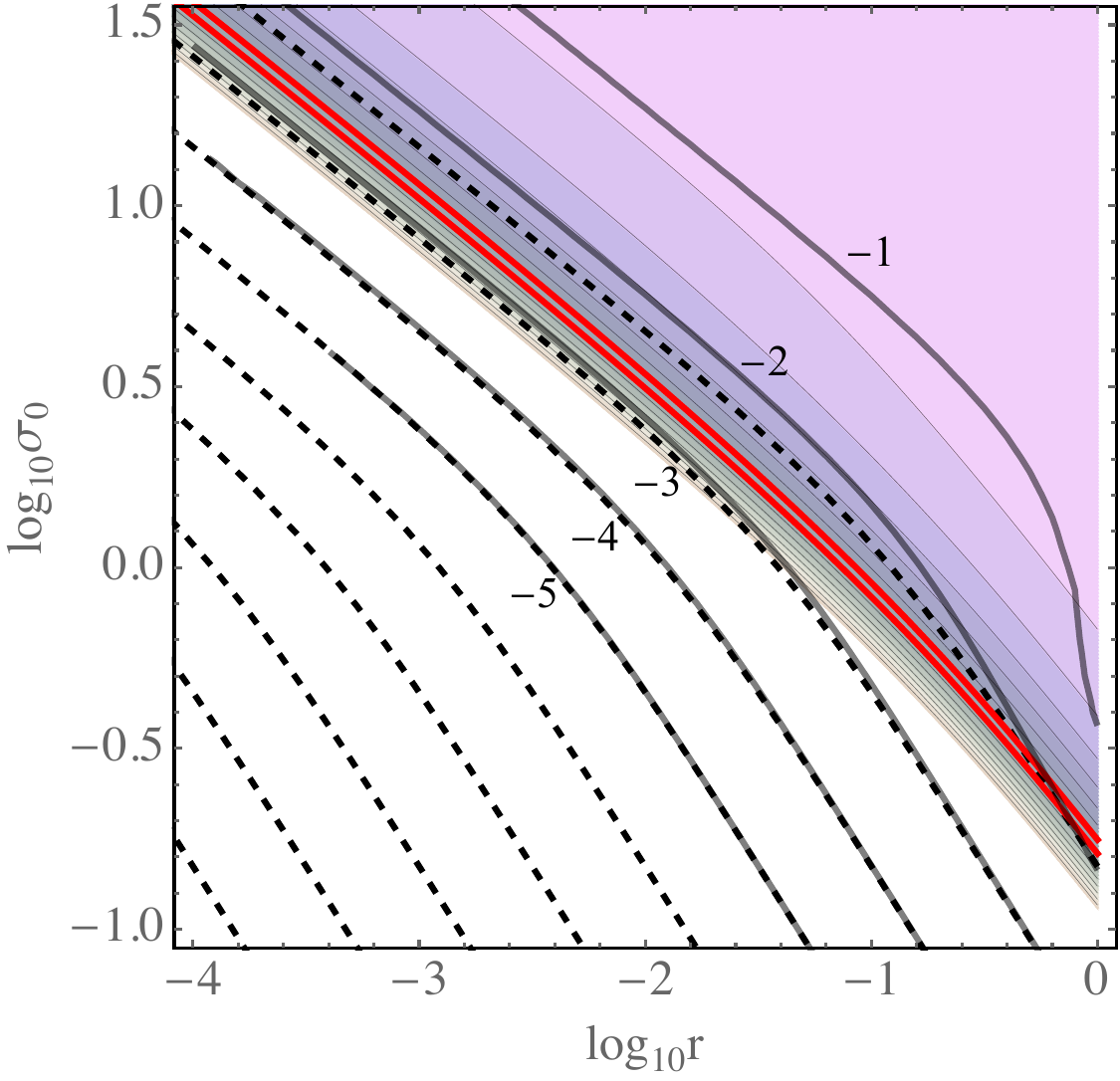}
\includegraphics[width=0.05\textwidth]{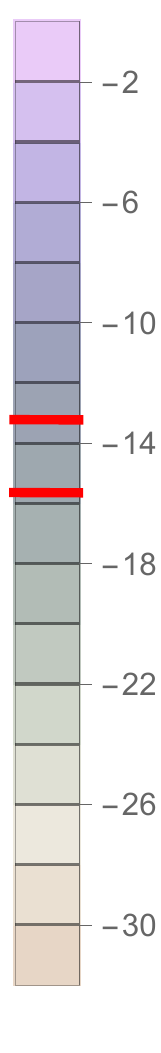}
\caption{The equal-$\sigma_\zeta^2$ contours (thick gray curves) and the equal-$\beta_\text{tot}$ contours (in color scales).
The analytical approximation, Eq.~\eqref{Pzetaapprox}, 
for the equal-$\sigma_\zeta^2$ contours are also shown with dashed curves. 
The mass window where PBH can be all the dark matter ($M_\text{PBH}=6.4\times10^{-17}M_\odot\sim3.9\times10^{-12}M_\odot$) is 
depicted as a thin strip bounded by two thick red lines, where the analytical approximation fits the numerical result well.}\label{f:2contour}
\end{center}
\end{figure}

The PBH abundance we observe today is parameterized by $f_\text{PBH}(M)\equiv\Omega_\text{PBH}(M)/\Omega_\text{CDM}$, which is connected to $\beta(M)$ by \cite{Carr:2020gox}
\be
f_\text{PBH}^\text{tot}\approx1.65\times10^8\left(M_*/M_\odot\right)^{-1/2}\beta_\text{tot},
\ee
where $\beta_\text{tot}=\int\beta(M)(M/M_*)^{-1/2}\dif\ln M$.
The equal-$\beta_\text{tot}$ contours in the $(r,\sigma_0)$-plane are shown in Fig.~\ref{f:2contour}.
For $r\lesssim0.1$, $\beta_\text{tot}$ depends only on the combination $\sigma_0\sqrt r$, which can be fitted by
\be\label{betafit}
\beta_\text{tot}\approx\frac{1}{2}\text{erfc}\left(\frac{1.97}{\sigma_0\sqrt r}\right).
\ee


The smoothed variance of $\zeta$ can be calculated by the PDF of $\delta$,
\begin{align}\label{def:Pzeta}
\sigma_{\zeta}^2(R)=\int\zeta_\text{b}^2P_\delta[\delta;R]\dif\delta-\left(\int\zeta_\text{b}P_\delta[\delta;R]\dif\delta\right)^2,
\end{align}
where $\zeta_\text{b}$ is the bare curvature perturbation defined in \eqref{sol:zeta} as a function of $\delta$.
The total power $\sigma_\zeta^2\equiv\int\mathcal P_\zeta(k)\dif\ln k$ is 
given by the $R\to0$ limit of \eqref{def:Pzeta}.
When $\zeta_\text{b}\lesssim1$, $\zeta_\text{b}$ can be approximated by \eqref{zetaseries},
\be\label{zetaLO}
\zeta_\text{b}\approx\frac r3\left(2\delta+\delta^2\right),
\ee
which is in the form of a quadratic local non-Gaussianity. Substituting  \eqref{zetaLO} into \eqref{def:Pzeta}, we have
\be\label{Pzetaapprox}
\sigma_\zeta^2
\approx\frac{4}{9}r^2\sigma_0^2\left(1+\frac12\sigma_0^2\right).
\ee
\eqref{Pzetaapprox} fits the numerical result quite well when $\sigma_\zeta^2\lesssim10^{-2}$, shown in Fig. \ref{f:2contour}. 
As $\sigma_\zeta^2$ and $\beta_\text{tot}$ depends only on $\sigma_0\sqrt r$ when $r\lesssim0.1$ and $\sigma_0\gtrsim1$, their contours are parallel to each other and the PBH abundance only depends on the amplitude of the power spectrum $\sigma_\zeta^2$ there.

In the vicinity of $r=1$, we have $\zeta_\text{b}\approx(2/3)\ln\left|1+\delta\right|$ from \eqref{zetaeqn}. Interestingly, this logarithmic form is the same as the fully nonlinear curvature perturbation generated in ultra-slow-roll inflation~\cite{Cai:2018dkf,Atal:2019cdz,Atal:2019erb}, from which the PBH formation has been calculated by \eqref{Pchain} in Ref.\cite{Biagetti:2021eep}. 
In our model, the full nonlinear region ($r\sim1$ and $\sigma_0\gtrsim1$) produces too much PBHs and is of no cosmological interest. 

\begin{figure}[htbp]
\begin{center}
\includegraphics[width=0.48\textwidth]{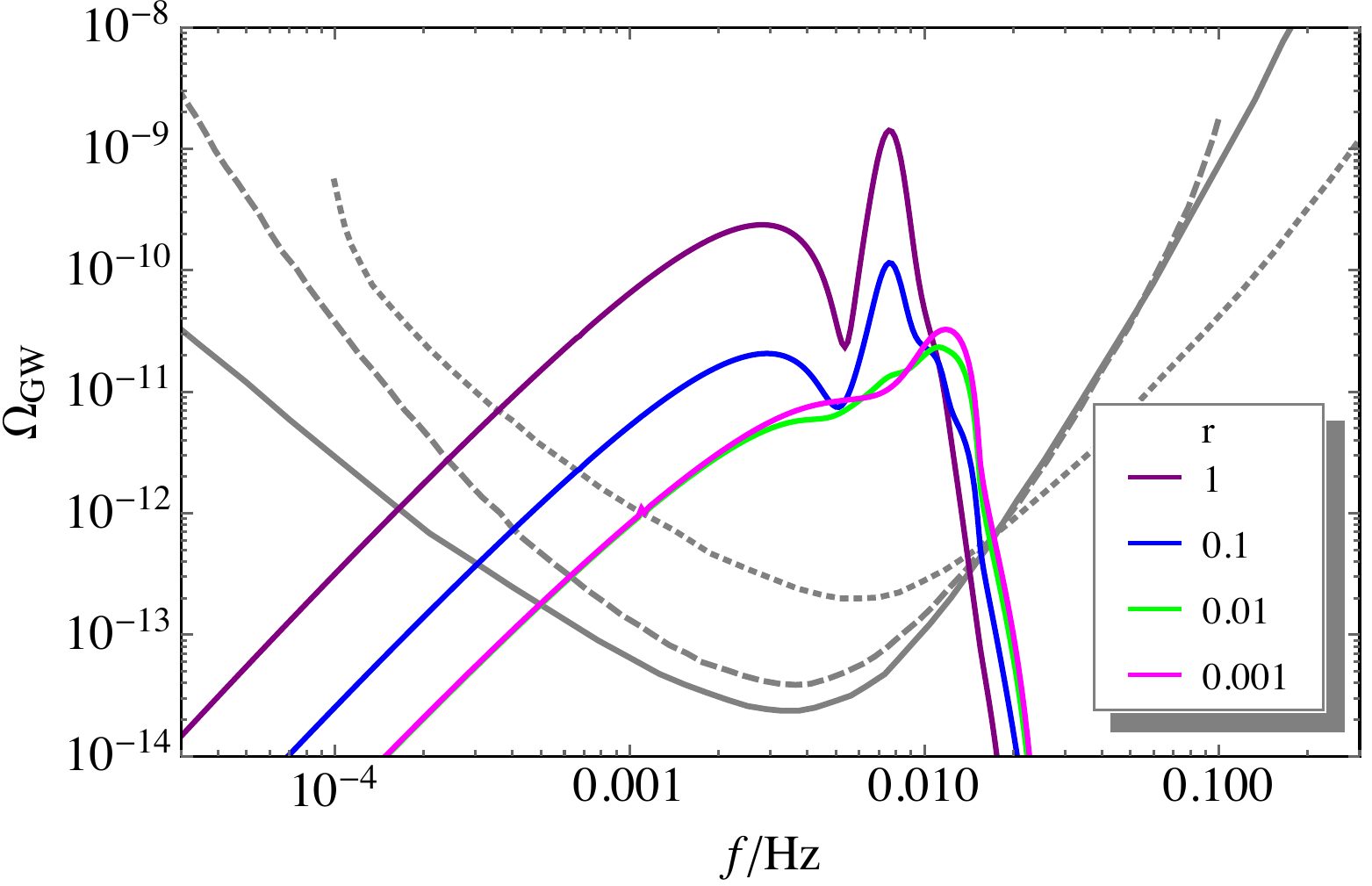}
\caption{The energy spectrum of the GWs induced by the curvature perturbation in the non-minimal curvaton model for several values of the curvaton fraction $r$ at its decay time.
We fix $f_\text{PBH}^\text{total}=1$ at $M_*=10^{-12}M_\odot$, corresponding to $f_*=6.7\times10^{-3}~\text{Hz}$. In the small-$r$ limit, the peak value levels off to a constant, $2.95\times10^{-11}$. The gray curves are the power-law integrated sensitivity of LISA (solid) \cite{Bartolo:2016ami}, Taiji (dashed)~\cite{Wang:2021njt}, and TianQin (dotted)~\cite{Liang:2021bde}.}\label{f:IGW}
\end{center}
\end{figure}

If the PBHs do not overclose our universe, i.e., if the energy density of PBHs is smaller than the critical density today, $\zeta$ can be well approximated by the quadratic local non-Gaussian form $\zeta=\zeta_g+F_\text{NL}(\zeta_g^2-\langle\zeta_g^2\rangle)$, with $\zeta_g=(2/3)r\delta$ and $F_\text{NL}=3/(4r)$.
This implies that the power spectrum of $\zeta$ consists of the Gaussian component and the non-Gaussian component given by convolution of two Gaussian spectra. 
The power spectrum $\calP_\delta(k)$ determines the Gaussian component of the curvature perturbation spectrum,
$\calP_{\zeta_g}(k)=(4/9)r^2\calP_\delta(k)$, while the non-Gaussian component is the convolution of $\calP_{\zeta_g}$. 
The induced GWs can be calculated by following the computations for the quadratic local non-Gaussianity case~\cite{Cai:2018dig,Unal:2018yaa,Adshead:2021hnm}. 
The resultant GW spectrum is shown in Fig.~\ref{f:IGW} for $f_\text{PBH}^\text{tot}=1$ at $M_*=10^{-12}M_\odot$, with the width $\Sigma=0.1$. 
The peak amplitude of the induced GW is roughly $0.1\Omega_\text{r}$ times the square of the curvature perturbation spectrum at the peak~\cite{Pi:2020otn,Domenech:2021ztg}. As is shown in Fig.~\ref{f:2contour}, if the PBH abundance is fixed, $\sigma_\zeta^2$ as well as $\Omega_\text{GW}$ decreases as $r$ decreases, but levels off to a constant as $r\to0$. This is because both $\sigma_\zeta^2$ and $\beta_\text{tot}$ depend only on the combination $\sigma_0\sqrt r$ in this limit. 
In the asteroid-mass window, PBHs can be all the dark matter, which requires $\sigma_0\sqrt r\sim0.36$ from \eqref{betafit}. 
The associated induced GW spectrum reaches its maximum,
\be\label{OmegaGW}
\Omega_\text{GW}\sim10^{-6}\sigma_\zeta^4\sim10^{-6}\times\frac{4}{81}\left(\sigma_0\sqrt r\right)^8\sim10^{-11},
\ee
at $\sim0.03$ Hz. This is the lower bound of $\Omega_\text{GW}$ when $f_\text{PBH}^\text{total}=1$. 
As is shown in Fig.~\ref{f:IGW}, \eqref{OmegaGW} is well above the power-law integrated sensitivity curves of LISA, Taiji, and TianQin around $10^{-2}$ Hz. 
So we reach the conclusion that the induced GWs must be detectable by LISA (and/or other planned space interferometeric observatories like Taiji/TianQin/BBO/DECIGO), if PBHs formed from the peaked curvature perturbation in the curvaton scenario constitute all the dark matter. This conclusion is in complete agreement with \cite{Cai:2018dig}, where the quadratic local non-Gaussianity is assumed {\it a priori} rather than derived. Here we have shown that the non-Gaussianity in the curvaton scenario can be well described by the quadratic form as long as we focus on the parameters of cosmological interest. 

\textit{Conclusion.}---Based on the curvaton scenario with a non-trivial field metric, we constructed a mechanism to enhance the curvature perturbation on small scales. The resulting curvature perturbation is fully non-Gaussian, while we use the probability distribution function to calculate the PBH formation with all the nonlinear effects taken into account. The PBH formation in curvaton scenario based on the expansion series of the curvature perturbation up to quadratic order has been studied before~\cite{Kawasaki:2012wr,Firouzjahi:2012iz,Kohri:2012yw,Bugaev:2013vba,Young:2013oia,Ando:2017veq,Ando:2018nge,Chen:2019zza,Inomata:2020xad,Liu:2021rgq}, but our paper is the first to study the PBH formation for the fully nonliear curvature perturbation.

Once we focus on the parameter space of cosmological interest, amazingly the curvature perturbation can be well approximated by a quadratic local non-Gaussianity $\zeta=\zeta_g+F_\text{NL}(\zeta_g^2-\langle\zeta_g^2\rangle)$. In this form the nonlinear parameter $F_\text{NL}=3/(4r)=\Omega_{\chi,\text{dec}}^{-1}-1/4$ can freely go to very large values, which is impossible in an expansion series. We further calculate the energy spectrum of the concomitant induced GWs under this approximation. Due to the enhancement of the PBH abundance by the non-Gaussianity, the power spectrum of the curvature perturbation required to generate a fixed amount of PBHs is suppressed when the non-Gaussianity is large. Yet the suppression has a lower bound when $F_\text{NL}\to\infty$. For PBHs to be all the dark matter, 
we have $\sigma_\zeta^2\gtrsim2\times10^{-3}$ and $\Omega_\text{GW}\gtrsim3\times10^{-11}$, 
which must be detectable by the planned interferometeric GW detectors in space.


In the region of $r\sim1$ and $\delta\gtrsim1$, the quadratic approximation breaks down and all the higher order terms should be taken into account. They sum up to be the similar logarithmic form to that in the ultra-slow-roll inflation ~\cite{Atal:2019cdz,Atal:2019erb,Kitajima:2021fpq,Biagetti:2021eep,Escriva:2022pnz,Pi:2022ysn}, as $\zeta=(2/3)\ln|1+\delta|$. Assuming the PDF of $\delta$ is Gaussian, this gives a non-Gaussian exponential tail of the PDF for the PBH formation. We leave the connection of these two models for future work.



\vspace{1em}
\begin{acknowledgements}
\textit{Acknowledgement.}---The work of S.P. is supported by the National Key Research and Development Program
of China Grant No. 2021YFC2203004, 
by Project 12047503 of the National Natural Science Foundation of China, by JSPS Grant-in-Aid for Early-Career Scientists No. JP20K14461, 
and in part by the International Centre for Theoretical Sciences (ICTS) for the online program -- ICTS Summer School on Gravitational-Wave Astronomy (code: ICTS/gws2021/7).
This work is also supported in part by the JSPS KAKENHI
Nos. JP19H01895, JP20H04727, and JP20H05853,
and in part by the World Premier International Research Center Initiative (WPI Initiative), MEXT, Japan. 
\end{acknowledgements}

\bibliography{ref}  

\end{document}